\newcommand{\green}[1]{ {\textcolor{olive}{#1}} }
\newcommand{\red}[1]{ {\textcolor{red}{#1}} }
\newcommand{\blue}[1]{ {\textcolor{blue}{#1}} }
\newcommand{\cyan}[1]{ {\textcolor{cyan}{#1}} }
\documentclass[conference]{IEEEtran}

\newcommand{\ab}[1]{{\bf [\color{teal} AB: {#1}]}}

\ifCLASSINFOpdf
\else
\fi

\usepackage{amsbsy,amscd,amsfonts,amsgen,amsmath,amsopn,amssymb,amstext,amsthm,amsxtra}
\usepackage{graphicx}
\usepackage{subfigure}
\usepackage{verbatim}
\usepackage{citesort}
\usepackage{url}
\usepackage{color}
\usepackage{comment}
\usepackage{tikz}
\usepackage{float}

\theoremstyle{definition}

\newcommand{\x}{\mathbf{x}}
\newcommand{\y}{\mathbf{y}}
\newcommand{\A}{\mathbf{A}}

\newcommand{\X}{\mathbf{X}}

\newcommand{\z}{\mathbf{z}}

\usepackage[margin=1in,footskip=0.25in]{geometry}
\DeclareMathOperator*{\argmin}{argmin}
\begin{document}
\title{
\vspace*{3mm}
{Nonlinear Function Estimation\\ 
with Empirical Bayes\\ 
and Approximate Message Passing}
}

\author{\IEEEauthorblockN{Hangjin Liu,$^\dagger$ You (Joe) Zhou,$^\dagger$ Ahmad Beirami,$^\ddagger$ and Dror Baron$^\dagger$}
\IEEEauthorblockA{$^\dagger$Department of Electrical and Computer Engineering, North Carolina State University, Raleigh, NC 27695, USA\\
$^\ddagger$Research Laboratory of Electronics, Massachusetts Institute of Technology, Cambridge, MA 02139, USA
\\
Email: \{hliu25,yzhou26,barondror\}@ncsu.edu, beirami@mit.edu}}
%\IEEEauthorblockA{}
%\IEEEauthorblockA{}

%Telephone: (800) 555--1212, Fax: (888) 555--1212}
\maketitle

\begin{abstract}
Nonlinear function estimation is core to modern machine learning applications.
In this paper, to perform nonlinear function estimation, we reduce a nonlinear
inverse problem to a linear one using a polynomial kernel expansion.
These kernels increase the feature set, and
may result in poorly conditioned matrices.
Nonetheless, we show several examples where
the matrix in our linear inverse problem 
contains only mild linear correlations among columns.
The coefficients vector is modeled within a Bayesian setting for which 
approximate message passing (AMP),
an algorithmic framework  for  signal  reconstruction,
offers Bayes-optimal signal reconstruction quality.
While the Bayesian setting limits the scope of our work, it is a first step 
toward estimation of real world nonlinear functions.
The coefficients vector is estimated using two AMP-based approaches,
a Bayesian one and empirical Bayes.
Numerical results confirm that our AMP-based approaches learn the function
better than LASSO, offering markedly lower error in predicting test data.
\end{abstract}



\IEEEpeerreviewmaketitle

%------------
\section{Introduction}
%------------

A pervasive trend in modern society
is that ever-larger
amounts of data are being collected and analyzed in order 
to explain various phenomena. In supervised learning, many variables 
(also referred to as features) that may
relate to and thus help explain the phenomena of interest are observed,
and the goal is to learn a function ---
often a nonlinear one --- that relates
the explanatory variables to the phenomena of interest. 
More specifically, we have a multivariate nonlinear function,
$\mathbf{f}(\cdot)$, and collect noisy samples of it;
our goal is to estimate $\mathbf{f}(\cdot)$.
At its core, this is {\em multivariate nonlinear function estimation};
it could also be interpreted as nonlinear regression or feature selection. 
Algorithms for solving such problems must 
be robust to noisy observations and outliers,
backed up by fundamental mathematical analysis,
support missing data, and have a fast implementation that
scales well to large-scale problems.
Such algorithms will impact many disciplines, such as health
informatics~\cite{Dalton2012}, %,Hua2005,Zollanvari2012,Dalton2013},
social networks, and finance~\cite{GrinoldKahn,FamaFrench1993,Jegadeesh1993}. %,Markowitz1952,Gyorfi2006}.

{\bf Example applications:}\
Let us describe how nonlinear function estimation can be used in
financial prediction~\cite{GrinoldKahn,FamaFrench1993,Jegadeesh1993}. 
A typical approach to estimate expected returns
uses a linear factor model, which is tuned to work
well on training data, 
$y_m = \sum_n {{\bf X}_{mn}} \theta_n + z_m$,
where $y_m$ is the price change of asset~$m$,
${{\bf X}_{mn}}$ is the exposure of 
asset~$m$ to factor $n$,
$\theta_n$ are the returns of factor~$n$, 
and $z_m$ is noise in asset~$m$.
We can express the linear model
in matrix vector form,
${\bf y} = {\bf X} {\boldsymbol{\theta}} + {\bf z}$,
where ${\bf X}$ is an input data matrix, 
by assigning $y_m$ as the $m$-th entry of the vector
${\bf y}$, $\theta_n$ as the 
$n$-th entry of the vector ${\boldsymbol \theta}$,
and $\X_{mn}$ as the 
element of the matrix ${\bf X}$ in row~$m$ and column~$n$ 
%\textcolor{green}{Dror points out to Hangjin that these aren't locations of samples in finance}
% \remove{With a slight abuse of notation, we use the same notation ${\bf X}$ both for locations where it's 
%\textcolor{green}{Dror to Hangjin - do you realize that it's is an informal word, and only a third grader would use it in written language? (Texting doesn't count!)}
%sampled and for organizing them in a data structure interchangeably.} 
The goal is to estimate ${\boldsymbol \theta}$ from ${\bf y}$,
${\bf X}$, and possible statistical knowledge about
${\boldsymbol \theta}$ and ${\bf z}$.  
We can see that financial prediction based on linear
models relies on solving linear inverse problems.
That said,  some factors relate
to returns in a nonlinear way~\cite{Chan1999stock},
and financial prediction could be improved using
nonlinear schemes.
%Similar approaches could also impact areas such as health
%informatics~\cite{Dalton2012} and %,Hua2005,Zollanvari2012,Dalton2013} and
%social networks.

Nonlinear modeling can also be used in health
informatics~\cite{Dalton2012}, %,Hua2005,Zollanvari2012,Dalton2013},
where $\y$ could measure patients' medical condition,
${\bf X}$ contains nonlinear exposure terms,
and ${\boldsymbol \theta}$ are explanatory variables that
drive the patients' condition. Our goal is to understand the
relationships between explanatory variables and patients' medical condition.

{\bf Main idea and contributions:}\
In this paper, as a first step toward learning nonlinear functions, 
we cast them as linear inverse problems using
{\em polynomial kernels}~\cite{Fan1995,Kim2002}
(Sec.~\ref{subsec:nonlinear}).
Incorporating kernels into the matrix maps the
nonlinear signal estimation procedure into a linear inverse problem 
but with an increased feature set, where the features are no longer 
\textit{independent and identically distributed} (i.i.d.).
Unfortunately, the kernels may create poorly conditioned matrices, and
many solvers for
linear inverse problems struggle with such matrices.
Nonetheless, the matrices in our linear inverse problems 
often contain only mild linear correlations among columns,
and are reasonably well conditioned.

While the polynomial kernels greatly increase the richness of the model class that 
captures the phenomena of interest, they also significantly increase the dimensionality of
the features. For example, $N$ factors evaluated with quadratic kernels will become
approximately $\frac{1}{2}N^2$ new factors. This large scale and well-conditioned
linear inverse problem is well-suited to {\em approximate message passing}
(AMP)~\cite{DMM2009,Montanari2012}, an algorithmic framework for
signal reconstruction that is asymptotically optimal for large scale
linear inverse problems in the sense that it achieves best-possible reconstruction
quality~\cite{GuoBaronShamai2009,RFG2009}. Our AMP-based 
algorithms improve reconstruction
quality of the coefficients vector ${\boldsymbol \theta}$, leading to better estimation of
the nonlinear function.

Two AMP-based approaches are considered. The first 
follows a {\em Bayesian framework}, where we
assume that the coefficients vector follows some known probabilistic
structure. While the Bayesian framework is naive and limited in scope,
our past work has shown that universal approaches that adapt to 
unknown statistical distributions can be integrated within solvers 
for linear inverse problems~\cite{BaronDuarteAllerton2011,ZhuBaronDuarte2014_SLAM,MaZhuBaron2016TSP},
thus bypassing the Bayesian limitation. The linear inverse problems 
resulting from our polynomial kernel expansion
is solved using an AMP-based algorithm,
whose Bayes optimality ensures that our function estimation procedure can 
succeed despite using fewer and noisier samples than other methods. 

The second AMP-based approach uses {\em empirical Bayes}~\cite{MTKB2014ITA},
where the coefficients vector is assumed to follow some parametric distribution,
and in each iteration of AMP we plug maximum likelihood parameter estimates 
into a parametric Bayesian denoiser.

The resulting algorithms will allow data to better model
dependencies between explanatory variables and
phenomena of interest. These algorithms could also help
reconstruct signals acquired by nonlinear analog
systems, allowing hardware designers
to exploit nonlinearities rather than avoid
them. 

{\bf Organization:}\
The rest of the paper is organized as follows.
Section~\ref{sec:background} provides background content.
Details of our approach for estimating multivariate nonlinear functions
appear in Section~\ref{subsec:nonlinear}.
Numerical results appear in Section~\ref{sec:numerical}, and 
Section~\ref{sec:conclusions} concludes.

%----
\section{Background}
\label{sec:background}
%---

%----
\subsection{Inverse problems}
%---

We present a flexible formulation for nonlinear function estimation in the form of a
nonlinear {\em inverse problem}.
We observe $M$ independent samples of the form 
$\{(\mathbf{x}_{m}, y_{m})\}_{{m} \in {\{1,\ldots, M\}}}$, 
where $(\mathbf{x}_{m}, y_{m}) \in \mathbb{R}^N \times \mathbb{R}$, 
through a nonlinear function 
$f(\cdot)$ and additive noise,
\begin{equation}
\label{eq:y_m}
    y_{m} = f({\bf x}_{m}) + z_{m},
\end{equation}
for all ${m} \in \{1,\ldots,M\}$. In other words,
the input data matrix ${\bf X} {\in\mathbb{R}^{M\times N}}$, 
where ${\bf X}$ are locations of samples,
will be processed by applying a multivariate operator,
${\bf f}(\cdot):{\mathbb{R}^M\times}N\rightarrow {\mathbb{R}^M}$, 
such that $\bf f$ applies $f$ on each individual row
of the data matrix $\bf x$,
with additive noise, ${\bf z} \in {\mathbb{R}^M}$,
resulting in noisy measurements, 
\begin{equation}
\label{eq:nonlinear_inverse}
{\bf y} = {\bf f}({\bf X})+{\bf z}\in \mathbb{R}^M.
\end{equation}
While the reader is likely familiar with linear inverse problems, 
where the operator ${\bf f}$ boils down to multiplication by a
coefficients vector $\boldsymbol{\theta}$, i.e., 
${\bf f}({\bf X}) = {\bf X} \boldsymbol{\theta}$,
our main interest is in nonlinear inverse problems.

We highlight that many ``rules of thumb" that the
sparse signal processing community has claimed,
for example that sparse signals can be reconstructed from a small
number of linear measurements, $M<N$, may break down when the
measurement noise ${\bf z}$ is large or the operator
${\bf f}(\cdot)$ contains significant nonlinearities.

%----
\subsection{Approximate message passing (AMP)}
\label{subsec:AMP}
%---

One approach for solving linear inverse problems
is AMP~\cite{DMM2009,Montanari2012},
which is an iterative algorithm that successively converts
the matrix problem to scalar channel denoising
problems with \textit{additive white Gaussian noise} (AWGN).
AMP is a fast approximation to precise message
passing (cf. Baron et al.~\cite{CSBP2010},
Montanari~\cite{Montanari2012}, and references therein),
and has received considerable attention because of its fast
convergence and the {\em state evolution} (SE)
formalism~\cite{DMM2009,Bayati2011,Montanari2012}, 
which characterizes how the \textit{mean squared error} (MSE)
achieved by the next iteration of AMP can be predicted using the
MSE performance of the denoiser being used. 
AMP solves the following linear inverse problem,
%\ab{I suggest changing {$\bf x$} to $\boldsymbol{\theta}$ in describing the inverse problem to make the notation consistent. Using {$\bf x$} could be very confusing as later the data will be embedded in the measurement matrix $\bf A$. In this case in fact ${\bf A} = {\bf X}$, and ${\bf f}({\bf X}) = {\bf X} \boldsymbol{\theta}$.}
\begin{equation}
{\bf y}={\bf X} \boldsymbol{\theta}+{\bf z},
\label{eq:linear_inverse}
\end{equation}
where the empirical {\em probability density function} (pdf) of  $\boldsymbol{\theta}$ follows $p_{\boldsymbol{\theta}}(\boldsymbol{\theta})$, the
operator~${\bf f(X)}$ multiplies ${\bf X}$ by 
the unknown coefficients vector $\boldsymbol{\theta}$,
and ${\bf z}$ is AWGN with variance $\sigma_Z^2$.
Although the AMP literature mainly considers i.i.d. 
Gaussian matrices, approaches such as damping~\cite{Rangan2014ISIT}
and Swept AMP~\cite{Manoel2015SweptAM} have been proposed to deal
with more general matrices.
After initializing ${\boldsymbol{\theta}}^0$ and ${\bf r}^0$,
AMP~\cite{DMM2009,Montanari2012}
proceeds iteratively according to
\begin{align}
\boldsymbol{\theta}^{t+1} &=& \eta^t({\bf X}^T{\bf r}^t+{{\boldsymbol{\theta}}}^t), \nonumber \\
{\bf r}^t &=& {\bf y}-{{\bf X}{\boldsymbol{\theta}}}^t+\frac{1}{R}{\bf r}^{t-1}
\langle\eta^{t-1'}({\bf X}^T{\bf r}^{t-1}+{{\boldsymbol{\theta}}}^{t-1})\rangle\label{eq:AMPiter},
\end{align}
where $(\cdot)^T$ denotes the transpose, 
\[
R=M/N
\]
is the {\em measurement rate}, $\eta^t(\cdot)$ is a {\em denoising function},
and~$\langle{\bf u}\rangle=\frac{1}{N}\sum_{i=1}^N u_i$
for some vector~${\bf u}\in\mathbb{R}^N$.
The denoising function~$\eta^t(\cdot)$ operates in a
symbol-by-symbol manner (also known as {\em separable})
in the original derivation of AMP~\cite{DMM2009,Montanari2012}.
That is,
$\eta^t({\bf u})=(\eta^t(u_1),\eta^t(u_2),...,\eta^t(u_N))$
and
$\eta^{t'}({\bf u})=(\eta^{t'}(u_1),\eta^{t'}(u_2),...,\eta^{t'}(u_N))$,
where $\eta^{t'}(\cdot)$ denotes the derivative of $\eta^t(\cdot)$.

A useful property of AMP 
in the {\em large system limit} ($N, M \rightarrow \infty$ 
with the measurement rate $R$ constant)
is that at each iteration,
the vector~${\bf X}^T{\bf r}^t+{\boldsymbol{\theta}}^t \in\mathbb{R}^N$
in (\ref{eq:AMPiter}) is equivalent to
the unknown coefficients vector
$\boldsymbol{\theta}$ corrupted by AWGN.
This property is based on the decoupling
principle~\cite{GuoVerdu2005,GuoBaronShamai2009,GuoWang2008},
which states that the posterior of a linear inverse problem
(\ref{eq:linear_inverse}) is statistically equivalent
to a scalar channel.
We denote the equivalent scalar channel at iteration~$t$ by
\begin{equation}
{\bf q}^t = {\bf X}^T{\bf r}^t+{\boldsymbol{\theta}}^t = \boldsymbol{\theta} + {\bf v}^t,
\label{eq:scalar_t}
\end{equation}
where $v^t_i\sim\mathcal{N}(0,\sigma_t^2)$,
and $\mathcal{N}(\mu,\sigma^2)$ is a Gaussian
pdf with mean~$\mu$ and variance $\sigma^2$.
AMP with separable denoisers, which are optimal for
i.i.d.\ signals, has been rigorously proved to obey
SE~\cite{Bayati2011}.
However, we will see in Section~\ref{subsec:UD} that non-i.i.d.\
signals can be denoised better using non-separable denoisers.

Another useful property of AMP in the large system limit involves
a Bayesian setting where a prior distribution for the 
coefficients vector $\boldsymbol{\theta}$ is available.
In such Bayesian settings, AMP can use denoiser functions $\eta^t(\cdot)$ that minimize 
the MSE in each iteration~$t$~\cite{Bayati2011}.
Using such MSE-optimal denoisers, the MSE performance of AMP~\eqref{eq:AMPiter} 
approaches the \textit{minimum mean squared error} (MMSE) as $t$ is increased.

\begin{figure*}[t]
\vspace*{3mm}
\begin{equation}
\label{eq:big_linear_system}
{\bf X}_Q = 
\begin{aligned}
    \begin{bmatrix}
    1  & x_{11} \cdots x_{1N} & x^2_{11} \dots x^2_{1N}   & x_{11}x_{12} \dots  x_{1(N-1)}x_{1N}\\
    1  & x_{21} \dots x_{2N} & x^2_{21} \dots x^2_{2N}   & x_{21}x_{22} \dots  x_{2(N-1)}x_{2N}\\
    \vdots & \vdots & \vdots &\vdots  \\
    1  & x_{M1} \dots x_{MN} & x^2_{M1} \dots x^2_{MN}   & x_{M1}x_{M2} \dots  x_{M(N-1)}x_{MN}
    \end{bmatrix}.
\end{aligned}
\end{equation}
\end{figure*}

%----
\subsection{Non-scalar denoisers}
\label{subsec:UD}
%---

While i.i.d. signals can be denoised in a scalar separable fashion
within AMP, where each signal entry is denoised using the same scalar denoiser,
real-world signals often contain dependencies between signal entries. 
For example, adjacent pixels in images are often similar in value,
and scalar separable denoisers ignore these dependencies.
Therefore, we apply non-separable denoisers to process
non-i.i.d.\ signals within AMP.
For example, if ${\boldsymbol{\theta}}$ is a time series containing
dependencies between adjacent entries, then
we can use a sliding window denoiser that
processes entry~{$n$} of ${\boldsymbol{\theta}}$ using information
from its neighbors~\cite{SW_Context2009,MaZhuBaron2016TSP}.

\begin{comment}
Sivaramakrishnan and Weissman~\cite{SW_Context2009}
proposed a {\em universal denoiser} that approaches
the mean square error (MSE) of optimal \textcolor{yellow}{Bayesian}
sliding window denoisers for stationary ergodic
signals, despite not knowing the input statistics.
The main idea underlying their
scheme is to use clustering~\cite{MacQueen1967kmeans}
to convert a complicated non-i.i.d.\ denoising problem to
i.i.d.\ problems, which are solved with
scalar denoisers. We have implemented their
approach~\cite{MaZhuBaronAllerton2014,MaZhuBaron2015submit},
and use a Gaussian mixture (GM) fitting
algorithm~\cite{FigueiredoJain2002} as part of
the scalar denoising.
Numerical results show that universal denoising within AMP (AMP-UD)
offers favorable reconstruction quality.
\end{comment}

We will see in Section~\ref{subsec:nonlinear}
that our signal reconstruction problem includes
several types of coefficients in ${\boldsymbol \theta}$, 
and we expect dependencies between coefficients. Therefore, non-scalar denoisers
will be used within AMP to process non-i.i.d.\ coefficients.

%----
\section{Learning nonlinear functions}
\label{subsec:nonlinear}
%----

Having reviewed relevant background material, we now recast
nonlinear inverse problems (\ref{eq:nonlinear_inverse})
as linear inverse problems (\ref{eq:linear_inverse}) using
polynomial kernels~\cite{Fan1995,Kim2002}, which replace our input data matrix
$\bf X$ with transformations of $\bf X$~\cite{Hastie2001}.  

Our nonlinear model (\ref{eq:nonlinear_inverse})
is motivated by the inadequacy of linear relationships
in some applications. One example involves bioinformatics, where
genetic factors involve multiplicative interactions
among genes~\cite{Kekatos2011}.
Another application involving financial
prediction~\cite{GrinoldKahn,FamaFrench1993,Jegadeesh1993}, %,Markowitz1952,Gyorfi2006},
where the research and development expenditures of a firm
correlate with future returns in a nonlinear way~\cite{Chan1999stock}.
Similar ideas have been widely used in the machine learning community
under the context of polynomial kernel learning~\cite{Fan1995,Kim2002}, and
the kernel trick has been introduced to linear inverse problems by
Qi and Hughes~\cite{QiHughes2011}. A related model that learns %aims at learning
interactions among variables is the multi-linear model~\cite{Nazer2010},
where columns that involve auto-interaction are removed
from the polynomial model.  %%%

%----
\subsection{Basis expansion}
\label{subsec:Taylor}
%----

%\ab{Notation inconsistency}
Recall that in our inverse problem, ${\bf y}={\bf f(X)}+{\bf z}$, 
we define measurement $m\in\{1,\ldots,M\}$ as
%\[
%y_m=[{{\bf f}(\x)}]_m+{z_m},
%\]
%\[
$y_m=f(\x_m)+{z_m}$ (\ref{eq:y_m}).
%\]
Linear inverse problems make use of models that are linear in the input factors; 
they are mathematically and algorithmically tractable, and can be interpreted as a 
first-order Taylor approximation to $f(\x)$~\cite{Hastie2001}. 
However, in many applications, the true function $f(\x)$ is far from linear in $\x$.

A basis function expansion replaces $\x$ with transformations of $\x$~\cite{Hastie2001}.  
For  $\ell \in \{1, 2, \ldots, L\}$, $f(\x)$ is expressed as in the 
linear basis expansion of ${\x}$: 
\[
f(\x)=\sum_{\ell=1}^{L}\theta_\ell g_{\ell}(\x).
\] 
This model is linear in the new variable $g_{\ell}(\x)$, and $\theta_\ell$ are the coefficients.
Basis expansions allow us to use a linear model to characterize and analyze nonlinear functions.

%----
\subsection{Polynomial regression}
%----

We form a polynomial regression problem by applying 
a Taylor expansion to the multivariate nonlinear function
$f(\cdot)$~\cite{Kekatos2011}. In polynomial regression, 
we add to the original columns of the measurement matrix  ${\bf X}_{Q}$,
which represent individual explanatory variables, extra columns that 
represent interactions among variables.  

Let us elaborate on the quadratic case. While we will provide details
of a matrix ${\bf X}_{Q}$, that supports a quadratic Taylor expansion~\eqref{eq:big_linear_system},
the reader should be able to employ this concept for cubic expansions and beyond.
For each measurement, we use a Taylor expansion of the $N$ factor variables:
\begin{equation}
\label{Taylor-poly}
\begin{aligned}
y_m=&\hspace{0.05in}\theta_1+\sum_{n=1}^{N}[{\boldsymbol\theta}_{2}]_{n}x_{mn}+\sum_{n=1}^{N}[{\boldsymbol\theta}_{3}]_{n}x^2_{mn}\\
&+\sum_{n_1=1}^{N}\sum_{n_2=n_1+1}^{N}[{\boldsymbol\theta}_{4}]_{n}x_{mn_1}x_{mn_2}, 
\end{aligned}
\end{equation}
where $\theta_1$ is a constant,
$\boldsymbol{\theta_2},\boldsymbol{\theta_3}\in\mathbb{R}^N$ 
are coefficient vectors for linear and quadratic terms, respectively, 
$\boldsymbol{\theta_4}\in\mathbb{R}^\frac{N(N-1)}{2}$ is a coefficient vector for cross terms, 
and the subscript $n$ in $[{\boldsymbol\theta}_{4}]_{n}$ depends on $n_1$ and $n_2$. 

Our quadratic Taylor approximation is a basis expansion,
where we have chosen $g(\x)$ as follows:
({\em i}) $g(\x)=1$ corresponds to a DC constant %\ab{Is $c =1$?};
({\em ii}) $N$ linear terms corresponding to the original data, $g(\x)=x_n$, $n\in\{1,\ldots,N\}$;
({\em iii}) $N$ quadratic terms corresponding to squares of individual linear terms, $g(\x)=(x_n)^2$; and
({\em iii}) $\frac{N(N-1)}{2}$ cross terms corresponding to products of pairs of linear terms,
$g(\x)=x_{n_1} x_{n_2}$, where $n_2>n_1$, $n_1,n_2\in\{1,\ldots,N\}$.
We assume that the features matrix, $\X$, is i.i.d.\ zero mean Gaussian for ease of analysis;
different types of $\X$ are left for future work.

The polynomial regression model is formulated 
as a linear inverse problem (\ref{eq:linear_inverse}) in matrix vector form,
\[%\begin{equation}
\y=\X_{Q}{\boldsymbol{\theta}}+\z=\X_{Q}\begin{bmatrix}
     \theta_{1} \\
     \boldsymbol{\theta_{2}}\\
     \boldsymbol{\theta_{3}}\\
     \boldsymbol{\theta_{4}}
    \end{bmatrix}+\z,
\]%\end{equation}
where ${\boldsymbol\theta}\in{\mathbb{R}^{L}}$ is the coefficient vector, and $L$ is evaluated 
below (\ref{eq:L}).
In our matrix $\X_Q$ (\ref{eq:big_linear_system}), 
each row is an instance or sample, and each column 
is an attribute or feature.

Our goal is to estimate the regression coefficients in the vector ${\boldsymbol\theta}$ from  
$\X_Q$ and $\y$. The measurement matrix $\X_Q\in{\mathbb{R}^{M\times L}}$ will include one DC column, 
$N$ linear term columns, $N$ quadratics (squared column),
and $\frac{N(N-1)}{2}$ cross terms. This matrix has the form 
(\ref{eq:big_linear_system}), and it can be seen that
\begin{equation}
\label{eq:L}
L=1+2N+\frac{N(N-1)}{2}.
\end{equation}

To solve this linear inverse problem 
using an AMP-based approach, we normalize each column of  $\X_Q$, $[\X_Q]_\ell$
to have unit norm, where $\ell \in \{1,2,\dots,L\}$, and denote this
normalized matrix by $\X_Q'$,
\begin{equation}
\label{eq:quadratic_to_linear}
\y= \X_Q {\boldsymbol{\theta}}+\z=\X_Q'{\boldsymbol{\theta}'}+\z,
\end{equation} 
where each entry of $[\X_{Q}]_\ell$ obeys 
\[
[\X_{Q}']_{\ell m}=\frac{[\X_{Q}]_{\ell m}}{||[\X_{Q}]_\ell||_2},
\]
and the regression coefficients satisfy
\begin{equation}
\label{coeffs_norm}
{{\theta}'}_\ell={{\theta}}_\ell|| [\X_{Q}]_\ell||_2.
\end{equation} 

%----
\subsection{SVD of normalized quadratic matrix $\X_Q'$}
\label{subsec:SVD}
%----

While the normalized matrix $\X_Q'$ converts our quadratic nonlinear inverse
problem into a linear one, it contains dependencies between linear and quadratic columns 
as well as between the linear and cross terms. Unfortunately, it is well known that
many solvers for linear inverse problems struggle with such matrices.

Surprisingly, our matrix (\ref{eq:big_linear_system}) works well within some
AMP-based approaches, as will be demonstrated by numerical results in
Section~\ref{sec:numerical}.
Why does our matrix perform well within AMP?
{\em Despite containing dependencies between
columns, these dependencies are nonlinear in nature, and linear correlations
between columns turn out to be mild}. In fact, a {\em singular value decomposition} (SVD) 
of $\X_Q'$ reveals that it is reasonably well-conditioned.
In particular, we have seen numerically that most of the {\em singular values} (SVs) 
seem to follow the semicircle law. That said, the first (largest) SV is 
larger than suggested by the semicircle law.

To see why the first SV, $\sigma_1$, is larger, 
recall that $\X_Q'$ is comprised of one DC column,
$N$ linear term columns, $N$ quadratic ones, and $\frac{N(N-1)}{2}$ cross term columns.
Because $\X_Q'$ has unit norm columns, entries of the DC column are $1/\sqrt{M}$, 
and so the sum of elements of the first column is $\sqrt{M}$.
The $N$ quadratic columns are non-negative, and because they too have unit norm, 
the average squared value is $1/M$, suggesting that the average is $\Theta(1/\sqrt{M})$.
The sums of elements of all $N$ linear and $\frac{N(N-1)}{2}$ cross term columns are near zero,
because these are zero mean Gaussian {\em random variables} (RVs),
and products of zero mean Gaussian RVs, respectively.
We see that the first SV, $\sigma_1$, corresponds to 
an all constant (or roughly all constant) column multiplied by a row
that contains significant non-zero entries corresponding to the DC column and 
$N$ quadratic columns, while row entries corresponding to linear and cross term columns 
are close to zero. 

Under some assumptions, we can estimate the amount of energy represented by
the first SV, $\sigma_1^2$. Suppose that the original linear columns
are Gaussian, $X \sim {\cal{N}}(0,1)$.
Under this assumption, the quadratic element
$\chi=X^2$ has a chi-squared distribution, where $E[\chi]=E[X^2]=1$
and $\text{var}[\chi]=2$. Therefore, $E[\chi^2]=E[\chi]^2+\text{var}(\chi)=3$. As
we will need to normalize individual entries of quadratic terms by roughly $\sqrt{3M}$,
 the average energy of the DC component of these columns is $1/3$.
Similarly, it can be shown that linear and cross term columns have average 
energy $1/M$ aligned with the first singular column vector. In summary, the energy in 
$\sigma_1^2$ is comprised of 
({\em i}) unit energy for the DC column;
({\em ii}) $N/M$ for the $N$ linear columns;
({\em iii}) $N/3$ for the $N$ quadratic ones; and
({\em iv}) $\frac{N(N-1)}{2M}$ for cross term columns.
Therefore, we predict the total energy in $\sigma_1$ to obey
\begin{equation}
\label{eq:sigma1}
\sigma^2_{1,pred}=1+N/3+\frac{N(N+1)}{2M}.
\end{equation}

Our analysis of the first singular value is inaccurate, because the first singular 
vector column is only roughly constant, and while computing the SVD
this column is modified in order to maximize the energy of the 
first rank-one component. Therefore, $\sigma^2_{1,pred}$ 
can be interpreted as a {\em lower bound} for $\sigma^2_1$.
That said, numerical experiments 
presented in Table~\ref{tbl:sv1} show that our prediction (\ref{eq:sigma1}) 
provides a reasonable approximation.
In the table, results for several $(M,N)$ pairs are provided.
For each pair, we average empirical values for $\sigma_1^2$, the energy in 
the first SV, over 20 matrices; these empirical averages are 
compared to the prediction (\ref{eq:sigma1}).
It can be seen that $\sigma_1^2$ is typically larger by 0.6--0.75;
seeing that unit norm columns in the normalized matrix $\X_Q'$ imply
that the average SV has unit energy, this extra energy 
seems plausible.

Finally, although we have focused on the normalized quadratic matrix,
$\X_Q'$, in further numerical work (not reported here) we evaluated a cubic matrix with normalized 
columns. It too has an SVD where $\sigma_1$ is larger while other
SVs seem to follow the semicircle law.

\begin{table}[h!]
  \begin{center}
    \caption{Empirical value of $\sigma^2_1$ compared to our prediction (\ref{eq:sigma1}).}
    \label{tbl:sv1}
%    \hline
    \begin{tabular}{|l|l|r|r|r|} 
%      \textbf{Value 1} & \textbf{Value 2} & \textbf{Value 3}\\
%      $\alpha$ & $\theta$ & $\gamma$ \\
%      \hline
%      1 & 1110.1 & a\\
%      2 & 10.1 & b\\
%      3 & 23.113231 & c\\
\hline
$M$ & $N$ & $L$ (\ref{eq:L}) & $\sigma^2_1$ (empirical) & $\sigma^2_{1,pred}$ (\ref{eq:sigma1}) \\
\hline
1000 & 10  & 66   & 4.99  & 4.39 \\
1500 & 15  & 136  & 6.72  & 6.08 \\
2000 & 20  & 231  & 8.41  & 7.77 \\
3000 & 20  & 231  & 8.35  & 7.74 \\
3000 & 30  & 496  & 11.84 & 11.16 \\
4000 & 40  & 861  & 15.23 & 14.54 \\
4500 & 50  & 1326 & 18.63 & 17.95 \\
5000 & 60  & 1891 & 22.09 & 21.37 \\
5500 & 70  & 2556 & 25.51 & 24.79 \\
5000 & 80  & 3321 & 29.06 & 28.31 \\
6000 & 80  & 3321 & 28.94 & 28.21 \\
8000 & 80  & 3321 & 28.78 & 28.07 \\
8000 & 90  & 4486 & 32.26 & 31.51 \\
6000 & 100 & 5151 & 35.93 & 35.18 \\
8000 & 100 & 5151 & 35.66 & 34.96 \\
      \hline
    \end{tabular}
  \end{center}
\end{table}

%----
\subsection{AMP-based algorithm}
%----

We solve our linear inverse problem (\ref{eq:quadratic_to_linear}) using AMP,
where two points should be highlighted.
First, our denoiser can incorporate the Bayesian
prior information. Specifically, we use conditional expectation denoisers
that minimize the MSE~\cite{Bayati2011}.
Second, owing to the structure of our matrix (Section~\ref{subsec:SVD}),
various AMP variants that promote convergence can be 
used~\cite{Manoel2015SweptAM,VAMP2017,Rangan2014ISIT}. That said, 
these variants all have their shortcomings, and possible divergence of AMP 
should be tracked carefully.

%------------
\section{Numerical Results}
\label{sec:numerical}
%------------

Our construction of the quadratic polynomial regression model
in Section~\ref{subsec:nonlinear}
results in a linear inverse problem (\ref{eq:linear_inverse})
whose solution forms an estimate of a
multivariate nonlinear function (\ref{eq:nonlinear_inverse})
that relates the explanatory variables to the phenomena of interest. 
This resulting linear inverse problem will now be solved 
by two AMP-based approaches, Bayesian AMP and empirical Bayes.

%----
\subsection{Bayesian AMP}
%---

{\bf Non-i.i.d.\ model for ${\boldsymbol{\theta}}$:}\
Our Bayesian approach considers
four groups of coefficients (\ref{Taylor-poly}),
where $\theta_1\in \mathbb{R}$,
$\boldsymbol{\theta_2},\boldsymbol{\theta_3}\in\mathbb{R}^N$, and
$\boldsymbol{\theta_4}\in\mathbb{R}^\frac{N(N-1)}{2}$
are the DC, linear, quadratic, and cross term coefficients, respectively.
We modeled each individual entry among these $L$ coefficients as
{\em Bernoulli Gaussian} (BG), where the Bernoulli part is a probability $p$
that the entry is nonzero, in which case its distribution is zero mean Gaussian 
with some variance. To be specific, 
({\em i}) our DC coefficent obeys $\theta_1\sim\mathcal{N}(0,10)$,
meaning that it is zero mean Gaussian with variance 10;
({\em ii}) each entry among the $N$ linear term coefficients satisfies $[\boldsymbol{\theta}_2]_n\sim0.2\mathcal{N}(0,1)+0.8\delta_0$, 
i.e., zero mean unit norm Gaussian with probability 0.2, else zero;
({\em iii}) the $N$ quadratic term coefficients obey
$[\boldsymbol{\theta}_3]_n\sim0.2\mathcal{N}(0,0.5)+0.8\delta_0$; and 
({\em iv}) for the $\frac{1}{2}N(N-1)$ cross term coefficients, 
$[\boldsymbol{\theta}_4]_n\sim0.03\mathcal{N}(0,0.1)+0.97\delta_0$. 
Although the four groups of coefficients have different distributions, 
all $L$ entries that follow this model are statistically independent.

{\bf Baseline LASSO algorithm:}\
The baseline algorithm used to solve (\ref{eq:quadratic_to_linear})
is the \textit{least absolute shrinkage and selection operator}
(LASSO)~\cite{Tibshirani1996}, which minimizes the sum of squared errors 
subject to a constraint on the $\ell_1$ norm of the coefficients~\cite{Tibshirani1996}. 
In our polynomial model, the LASSO estimator $\widehat{\boldsymbol{\theta}}$ is calculated in 
Lagrangian form: 
\begin{equation}
    \widehat{\boldsymbol{\theta}}=\frac{1}{2}\argmin_{\boldsymbol{\theta}}||\y-\X_Q\boldsymbol{\theta}||_2^2+
 \sum_{j=1}^4 \lambda_j\| \boldsymbol{\theta}_j \|_1,
\end{equation}
where $\lambda_1, \ldots, \lambda_4$ are tuning parameters.
In principle, we could perform grid search over all four parameters, 
$\lambda_1, \ldots, \lambda_4$,
but it is computationally intractable. Therefore, we report the performance 
obtained by setting all parameters to be equal, which reduces
the search space.

{\bf AMP-based approach:}\
As a proof of concept, we have designed a denoiser specifically for our non-i.i.d.\ model.
Because all $L$ entries that follow this model are statistically independent,
we used $L$ scalar denoisers. However, because individual entries among our
four groups of coefficients, $\theta_1\in \mathbb{R}$,
$\boldsymbol{\theta_2},\boldsymbol{\theta_3}\in\mathbb{R}^N$, and
$\boldsymbol{\theta_4}\in\mathbb{R}^\frac{N(N-1)}{2}$
follow different distributions, four different scalar denoisers were used.
Details of Bayesian denoisers for BG signals appear in~\cite{VilaSchniter2011}.

{\bf Signal generation:}\ We evaluate the performance of AMP in the Bayesian 
setting, which is a planted inference problem.
The experiment allows us to validate the suitability of AMP for the quadratic basis, 
e.g.~\eqref{eq:big_linear_system}.

We generated the feature matrix, $\X$, as i.i.d.\ Gaussian with dimension $N=100$.
These linear terms were then transformed into a quadratic form $\X_Q'$ with normalized
columns (\ref{eq:quadratic_to_linear}). The number of columns in the normalized
matrix was $L=5151$ (\ref{eq:L}), and the number of rows $M=5400$, 
Next, we created quadratic multivariate functions 
by generating ${\boldsymbol \theta}$ vectors following our non-i.i.d.\ model. 
The expected energy of each group of coefficients satisfies
$E_{DC}=10$, $E_{linear}=0.2 \times N=20$, $E_{quadratic}=0.2 \times N \times 0.5=10$, and $E_{cross}=0.03\times\frac{N^2-N}{2}\times0.1=14.85$. 
Finally, the measurement noise ${\bf z}$ was AWGN with variance $\sigma_Z^2=0.004$. 

\begin{figure*}[h]
\centering
\begin{tabular}{cccc}
\hspace*{-5mm}
\includegraphics[width=0.5\textwidth]{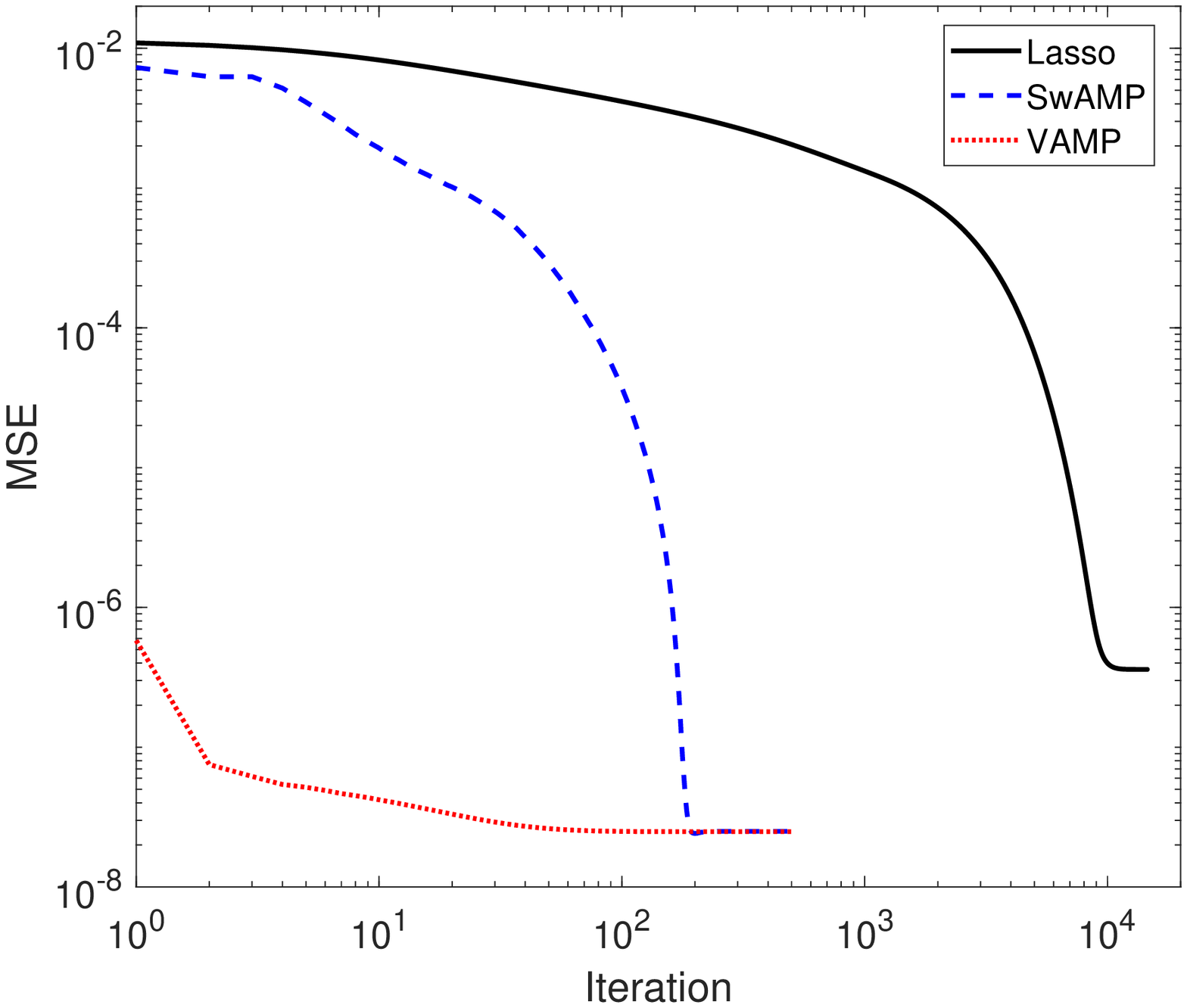} &
\includegraphics[width=0.5\textwidth]{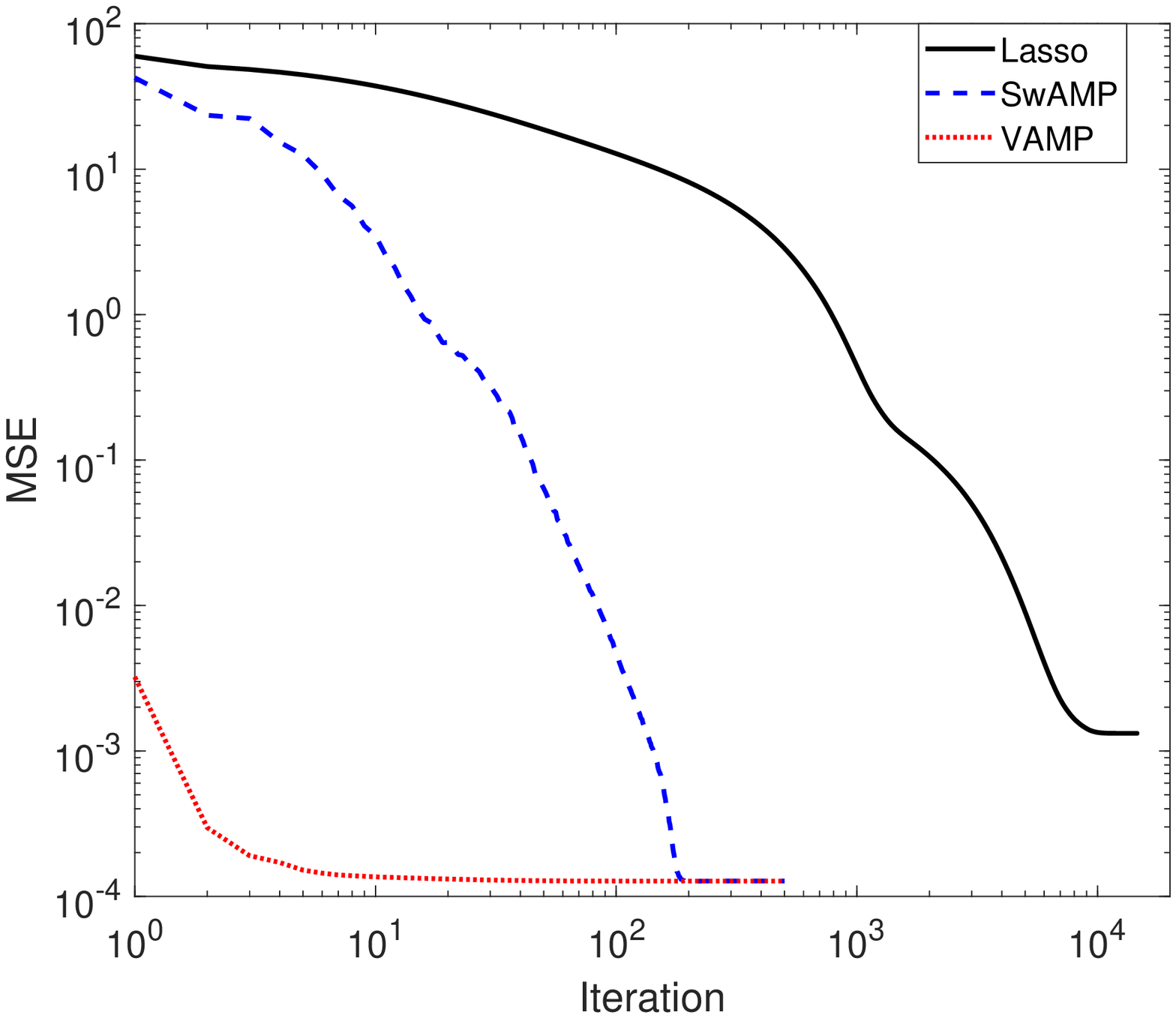} \\
\textbf{(a)}  & \textbf{(b)} \\[12pt]
\end{tabular}
\caption{ 
Performance of LASSO, SwAMP, and VAMP.
The MSE is shown in the vertical axis, while the horizontal axis reflects the iteration number, $t$.
Left panel {\bf (a):} MSE performance in recovering the unknown coefficients, $\boldsymbol{\theta}$.
Right panel {\bf (b):} MSE performance in predicting the test data.}
\label{fig:Bayes_model_Lasso_AMP}
\end{figure*}

{\bf MSE performance:}\ 
Fig.~\ref{fig:Bayes_model_Lasso_AMP} shows the MSE performance for estimated 
coefficients, ${\boldsymbol{\theta}}$. We estimated the coefficients using LASSO, 
{\em swept AMP} (SwAMP)~\cite{Manoel2015SweptAM} and \textit{vector AMP} (VAMP)~\cite{VAMP2017}.
The left panel of the figure shows the MSE obtained when estimating the original coefficients 
$\boldsymbol{\theta}$, 
where the estimator $\widehat{\bf{\theta}}$ can be calculated using~\eqref{coeffs_norm_est},
\begin{equation}
\label{coeffs_norm_est}
{\widehat{\bf{\theta}}}_\ell=\frac{\widehat{{\bf{\theta}}'}_\ell}{|| [\X_{Q}]_\ell||_2},
\end{equation}
$l\in\{1,\ldots,L\}$, and $\widehat{\boldsymbol{\theta}'}$ are estimated coefficients of 
${\boldsymbol{\theta}'}$.
SwAMP and VAMP both converge well for normalized quadratic matrices. 
However, it can be seen in Fig.~\ref{fig:Bayes_model_Lasso_AMP} that 
VAMP requires less than one hundred iterations to converge;
SwAMP requires a few hundred, and its individual iterations require 
more computation than those of VAMP;
our specific implementation of LASSO requires thousands of iterations. 
Because our AMP based approaches are expected to be Bayes optimal while LASSO
does not share these optimality properties, there is no surprise that 
AMP-based approaches obtain lower MSE.

To make sure that our function reflects the nonlinear function well,
the right panel of Fig.~\ref{fig:Bayes_model_Lasso_AMP} 
shows the MSE obtained when applying our estimated polynomial function to
predict test data,
\[\frac{||\y_{test}-\X_{test}\widehat{\boldsymbol\theta}||_2^2}{K}
=\frac{||\X_{test}({\boldsymbol\theta}-\widehat{\boldsymbol\theta})||_2^2}{K},\]
where we held back $K=600$ test measurements (recall that $M=5400$),
$\X_{test}\in\mathbb{R}^{K\times L}$ has the same
format as $\X_{Q}$, 
and $\y_{test}\in\mathbb{R}^{K}$. 
Note that the MSE for coefficients, $\boldsymbol{\theta}$, 
is inapplicable to real-world problems,
because the true coefficients do not exist, and we are merely modeling some 
nonlinear dependence as a low-order Taylor series. In our synthetic experiment,
we are using the MSE over the test data as a metric of interest.

%----
\subsection{Empirical Bayes}
%---

{\bf Nonlinear function:}\
Nonlinear function learning is now performed 
using empirical Bayes within AMP~\cite{MTKB2014ITA}.
We employ the quadratic formulation \eqref{eq:quadratic_to_linear} and learn the coefficients vector 
${\boldsymbol \theta}$ to approximate a family of (mildly) nonlinear functions,
\begin{equation}
\label{eq:nonfunc}
{\boldsymbol{y}}=\sum_{i=1}^{3} w_i \sin \left({\boldsymbol{X}} {\boldsymbol{\rho}_i}+{\boldsymbol{\phi}_i}\right)
+\bf z,
\end{equation}
where 
$w_1=0.1$, $w_2=0.3$, and $w_3=0.6$ are weights of the sinusoids,
${\boldsymbol{\rho}}_i \in \mathbb{R}^N$ is a BG vector,
${\boldsymbol{X}} {\boldsymbol{\rho}_i} \in \mathbb{R}^M$,
${\boldsymbol{\phi}} \in \mathbb{R}^M$ are phase shifts uniformly distributed
between 0 and $2\pi$,
the sine is applied element-by-element,
and the noise $\boldsymbol{z} \in \mathbb{R}^M$ is AWGN with variance $10^{-4}$.
Note that the vectors ${\boldsymbol{\rho}}_i$ are chosen to be sparse BG, 
in order for the coefficients vector $\boldsymbol{\theta}$ fit by AMP to
the quadratic expansion to also be sparse.

{\bf AMP-based empirical Bayes:}\ 
In contrast to the Bayesian case, we assume that 
$\boldsymbol{\theta}_2$, $\boldsymbol{\theta}_3$, and $\boldsymbol{\theta}_4$
are BG, and their parameters are estimated using {\em maximum likelihood} (ML) in each AMP iteration.
The DC coefficient $\theta_1$ is assumed to be Gaussian.
The ML parameters are plugged into Bayesian denoisers for the 4 components.

{\bf MSE performance:}\ 
We generated nonlinear functions and ran our empirical Bayes algorithm, LASSO,
and a pseudoinverse approach (least squares).
Each run of LASSO requires many iterations, and we use cross validation to regularize the 
parameter selection procedure.
AMP with damping requires fewer iterations than LASSO.
Empirical results for different measurement rates, $R=M/L$, appear in Table.~\ref{tab:MSE}. 
AMP obtains lower MSE than LASSO, which in turn obtains lower MSE than pseudoinverse.

\begin{table}[]
\centering
\caption{Empirical MSE on test data for nonlinear function estimation.}
\begin{tabular}{|c|c|c|c|}
\hline 
%\multirow
{\textbf{
\begin{tabular}[c]{@{}c@{}}Measurement \\Rate $R=\frac{M}{L}$\end{tabular}}} & \multicolumn{3}{c|}{
\textbf{\begin{tabular}[c]{@{}c@{}}Median MSE over\\20 Realizations\end{tabular}}}
\\ \cline{2-4} 
 & \textbf{LASSO} & \textbf{AMP} & \textbf{Pseudoinverse} \\ \hline
0.14 & 0.0382 & 0.0293 & 0.041 \\ \hline
0.28 & 0.0298 & 0.0228 & 0.033 \\ \hline
0.56 & 0.0063 & 0.0036 & 0.01 \\ \hline
\end{tabular}\label{tab:MSE}
\end{table}

%------------
\section{Discussion}
\label{sec:conclusions}
%------------

In this paper, we studied nonlinear function estimation, where a nonlinear function of interest
is regressed on a set of features. 
We linearized the problem by considering low-order polynomial kernel expansion, and solved the resulting linear inverse problem using {\em approximate message passing} (AMP). 
Numerical results confirm that our AMP-based approaches learn the function
better than the widely used 
{\em least absolute shrinkage and selection operator} (LASSO)~\cite{Tibshirani1996},
offering markedly lower error in 
predicting test data for both Bayesian and non-Bayesian settings.

While we have presented a first step toward estimating nonlinear functions
by appling AMP to polynomial regression, many open problems remain.

\begin{comment}
{\bf Fundamental performance limits:}\
We have described a practical algorithm that converts a nonlinear inverse
problem (\ref{eq:nonlinear_inverse}) into a linear inverse form (\ref{eq:linear_inverse}),
which is solved with AMP-based approaches. Although our numerical results have shown 
promising results for a \textcolor{yellow}{Bayesian} setting where the distribution of the coefficient
vector ${\boldsymbol \theta}$ was known, it is not clear whether this is any way optimal.
Although the \textcolor{yellow}{Bayesian} framework is often restrictive, it allows us to make rigorous
statements about its performance. A key question is whether our AMP-based approach
achieves fundamental estimation-theoretic limits in this \textcolor{yellow}{Bayesian} setting.
A more ambitious question is whether realistic function classes can be approximated
by Taylor series whose coefficients have properties for which AMP-based approaches
are optimal. Specifically, we can ask how many measurements are needed to reach 
desirable levels of function estimation quality.
Such questions are left for future work.
\end{comment}

{\bf Dependencies between coefficients:}\ 
In past work, we used non-scalar sliding window denoisers
to process coefficient vectors ${\boldsymbol \theta}$ that contained 
dependencies between entries~\cite{SW_Context2009,MaZhuBaron2016TSP}.
It is not clear whether similar dependencies will appear
in our ${\boldsymbol \theta}$. While it seems plausible
that exposure weights corresponding to the $N$ original columns,
the $N$ quadratic terms, and $N(N-1)/2$ cross terms will have different
distributions, it is not clear whether each group is i.i.d.
or contains intra-group dependencies.
In ongoing work, we are processing all
terms corresponding to the same original column
(the original column, its quadratic, and $N-1$ associated
product columns) together, which could be processed with
block denoising. This form of joint processing will support possible 
dependencies between lower order Taylor coefficients
and higher order ones; such dependencies have been 
noted between parent and children wavelet coefficients~\cite{Simoncelli1996}.

{\bf Other kernels:}
In this paper, we considered a second-order polynomial kernel. Future work will naturally extend to 
selecting the degree of the polynomial kernel as well. Further, we will consider other widely used kernels.
%such as the Gaussian kernel.

{\bf Results on real datasets:}\
While we reported promising results 
for nonlinear function estimation with AMP
in Bayesian and empirical Bayes settings, 
the performance of our algorithms must be tested on real datasets. 
In these datasets, various problems may appear, for example the prior
is unavailable;
the measurement matrix may be poorly conditioned; 
the function of interest may not belong to the hypothesis class;
and the noise may be heavy tailed~\cite{GrinoldKahn}, resulting in a 
mismatched estimation problem. We will explore the application of more advanced adaptive variants of 
AMP in the absence of a known prior~\cite{BaronDuarteAllerton2011,ZhuBaronDuarte2014_SLAM,MaZhuBaron2016TSP}.
When the true function does not belong to the hypothesis class, which are polynomials of 
degree two or three
in this paper, the best one can hope for is to recover the function of interest up to a 
projection error onto the hypothesis class. 
We will also explore the usual bias/variance trade-offs that arise in such settings.

{\bf Nonlinear acquisition and reconstruction:}\
Since the work of Gauss and his contemporaries~\cite{Gauss1809theoria},
hardware designers have been keenly aware that the mathematics
involved in processing linearly obtained measurements is more
mature than that for nonlinear measurements.
However, algorithms that estimate multivariate nonlinear
functions can also be used to reconstruct signals measured
nonlinearly. The same polynomial kernels~\cite{Fan1995,Kim2002}
used above to expand the matrix 
%${\bf A}$ to support nonlinear functions
can also be used to approximate a nonlinear function
with a linear one.
%${\bf A(\cdot)}$ with a linear inverse problem involving an expanded matrix
%${\bf \widetilde{A}}$ with far more columns.
Such advances will allow designers to stop worrying about
the nonlinearities inherent in many hardware systems.

\section*{Acknowledgment}
%------------

The authors are greatly indebted to Yanting Ma, who 
demonstrated favorable preliminary results for various AMP variants 
on a quadratic matrix. 
%AMP-MD was also inspired by Henry Pfister, who noted that
%while a prominent low rank component can interfere with other parts
%of the measurement matrix, it is also relatively easy to estimate 
%and thus correct for its impact on the matrix vector product.
DB also thanks Andrew Barron for helping him appreciate the
importance of nonlinear function estimation.
Finally, this work was partly supported by the National Science Foundation 
under Grant Nos. ECCS-1611112 and CNS 16-24770, and the industry members of the Center 
for Advanced Electronics in Machine Learning.

%--------------------------------------------------------------
% BIBLIOGRAPHY
%--------------------------------------------------------------
%\newpage
\bibliographystyle{IEEEtran}
\bibliography{cites}

\begin{thebibliography}{10}
\providecommand{\url}[1]{#1}
\csname url@samestyle\endcsname
\providecommand{\newblock}{\relax}
\providecommand{\bibinfo}[2]{#2}
\providecommand{\BIBentrySTDinterwordspacing}{\spaceskip=0pt\relax}
\providecommand{\BIBentryALTinterwordstretchfactor}{4}
\providecommand{\BIBentryALTinterwordspacing}{\spaceskip=\fontdimen2\font plus
\BIBentryALTinterwordstretchfactor\fontdimen3\font minus
  \fontdimen4\font\relax}
\providecommand{\BIBforeignlanguage}[2]{{%
\expandafter\ifx\csname l@#1\endcsname\relax
\typeout{** WARNING: IEEEtran.bst: No hyphenation pattern has been}%
\typeout{** loaded for the language `#1'. Using the pattern for}%
\typeout{** the default language instead.}%
\else
\language=\csname l@#1\endcsname
\fi
#2}}
\providecommand{\BIBdecl}{\relax}
\BIBdecl

\bibitem{Dalton2012}
L.~A. Dalton and E.~R. Dougherty, ``Optimal classifier with minimum expected
  error within a {B}ayesian framework - {P}art 1: {D}iscrete and {G}aussian
  model,'' \emph{Pattern Recognition}, vol.~46, pp. 1301--1314, Nov. 2012.

\bibitem{GrinoldKahn}
R.~C. Grinold and R.~N. Kahn, \emph{Active portfolio management: a quantitative
  approach for providing superior returns and controlling risk}.\hskip 1em plus
  0.5em minus 0.4em\relax McGraw-Hill Companies, 2000.

\bibitem{FamaFrench1993}
E.~Fama and K.~French, ``{Common risk factors in the returns on stocks and
  bonds},'' \emph{J. Finan. Econ.}, vol.~33, no.~1, pp. 3--56, 1993.

\bibitem{Jegadeesh1993}
N.~Jegadeesh and S.~Titman, ``{Returns to buying winners and selling losers:
  Implications for stock market efficiency},'' \emph{J. Finance}, vol.~48,
  no.~1, pp. 65--91, 1993.

\bibitem{Chan1999stock}
L.~K. Chan, J.~Lakonishok, and T.~Sougiannis, ``The stock market valuation of
  research and development expenditures,'' National Bureau of Economic
  Research, Tech. Rep., 1999.

\bibitem{Fan1995}
J.~Fan, N.~E. Heckman, and M.~P. Wand, ``Local polynomial kernel regression for
  generalized linear models and quasi-likelihood functions,'' \emph{J. Amer.
  Stat. Assoc.}, vol.~90, no. 429, pp. 141--150, 1995.

\bibitem{Kim2002}
K.~I. Kim, K.~Jung, and H.~J. Kim, ``Face recognition using kernel principal
  component analysis,'' \emph{IEEE Signal Process. Lett.}, vol.~9, no.~2, pp.
  40--42, 2002.

\bibitem{DMM2009}
D.~L. Donoho, A.~Maleki, and A.~Montanari, ``{Message passing algorithms for
  compressed sensing},'' \emph{Proc. Nat. Academy Sci.}, vol. 106, no.~45, pp.
  18\,914--18\,919, Nov. 2009.

\bibitem{Montanari2012}
A.~Montanari, ``Graphical models concepts in compressed sensing,''
  \emph{Compressed Sensing: Theory and Applications}, pp. 394--438, 2012.

\bibitem{GuoBaronShamai2009}
D.~Guo, D.~Baron, and S.~Shamai, ``A single-letter characterization of optimal
  noisy compressed sensing,'' in \emph{Proc. Allerton Conf. Commun., Control,
  and Comput.}, Sept. 2009, pp. 52--59.

\bibitem{RFG2009}
S.~Rangan, A.~K. Fletcher, and V.~K. Goyal, ``Asymptotic analysis of {MAP}
  estimation via the replica method and applications to compressed sensing,''
  \emph{CoRR}, vol. abs/0906.3234, June 2009.

\bibitem{BaronDuarteAllerton2011}
D.~Baron and M.~F. Duarte, ``Universal {MAP} estimation in compressed
  sensing,'' in \emph{Proc. Allerton Conf. Commun., Control, and Comput.},
  Sept. 2011, pp. 768--775.

\bibitem{ZhuBaronDuarte2014_SLAM}
J.~Zhu, D.~Baron, and M.~F. Duarte, ``Recovery from linear measurements with
  complexity-matching universal signal estimation,'' \emph{IEEE Trans. Signal
  Process.}, vol.~63, no.~6, pp. 1512--1527, Mar. 2015.

\bibitem{MaZhuBaron2016TSP}
Y.~Ma, J.~Zhu, and D.~Baron, ``Approximate message passing algorithm with
  universal denoising and {G}aussian mixture learning,'' \emph{IEEE Trans.
  Signal Process.}, vol.~65, no.~21, pp. 5611--5622, Nov. 2016.

\bibitem{MTKB2014ITA}
Y.~Ma, J.~Tan, N.~Krishnan, and D.~Baron, ``Empirical {B}ayes and full {B}ayes
  for signal estimation,'' \emph{Arxiv preprint arxiv:1405.2113v1}, May 2014.

\bibitem{CSBP2010}
D.~Baron, S.~Sarvotham, and R.~G. Baraniuk, ``Bayesian compressive sensing via
  belief propagation,'' \emph{IEEE Trans. Signal Process.}, vol.~58, no.~1, pp.
  269--280, Jan. 2010.

\bibitem{Bayati2011}
M.~Bayati and A.~Montanari, ``The dynamics of message passing on dense graphs,
  with applications to compressed sensing,'' \emph{IEEE Trans. Inf. Theory},
  vol.~57, no.~2, pp. 764--785, Feb. 2011.

\bibitem{Rangan2014ISIT}
S.~Rangan, P.~Schniter, and A.~Fletcher, ``On the convergence of approximate
  message passing with arbitrary matrices,'' in \emph{Proc. IEEE Int. Symp.
  Inf. Theory (ISIT)}, July 2014, pp. 236--240.

\bibitem{Manoel2015SweptAM}
A.~Manoel, F.~Krzakala, E.~W. Tramel, and L.~Zdeborov{\'a}, ``Swept approximate
  message passing for sparse estimation,'' in \emph{Proc. Int. Conf. Machine
  Learning}, vol.~37, July 2015, pp. 1123--1132.

\bibitem{GuoVerdu2005}
D.~Guo and S.~Verd{\'u}, ``Randomly spread {CDMA}: {A}symptotics via
  statistical physics,'' \emph{IEEE Trans. Inf. Theory}, vol.~51, no.~6, pp.
  1983--2010, June 2005.

\bibitem{GuoWang2008}
D.~Guo and C.~C. Wang, ``Multiuser detection of sparsely spread {CDMA},''
  \emph{IEEE J. Sel. Areas Commun.}, vol.~26, no.~3, pp. 421--431, Apr. 2008.

\bibitem{SW_Context2009}
K.~Sivaramakrishnan and T.~Weissman, ``A context quantization approach to
  universal denoising,'' \emph{IEEE Trans. Signal Process.}, vol.~57, no.~6,
  pp. 2110--2129, June 2009.

\bibitem{Hastie2001}
T.~Hastie, R.~Tibshirani, and J.~H. Friedman, \emph{The Elements of Statistical
  Learning}.\hskip 1em plus 0.5em minus 0.4em\relax Springer, Aug. 2001.

\bibitem{Kekatos2011}
V.~Kekatos and G.~Giannakis, ``Sparse {V}olterra and polynomial regression
  models: {R}ecoverability and estimation,'' \emph{IEEE Trans. Signal
  Process.}, vol.~59, no.~12, pp. 5907--5920, Dec. 2011.

\bibitem{QiHughes2011}
H.~Qi and S.~Hughes, ``{Using the kernel trick in compressive sensing: Accurate
  signal recovery from fewer measurements},'' in \emph{IEEE Int. Conf. Acoust.,
  Speech, Signal Process. (ICASSP)}.\hskip 1em plus 0.5em minus 0.4em\relax
  IEEE, 2011, pp. 3940--3943.

\bibitem{Nazer2010}
B.~Nazer and R.~Nowak, ``Sparse interactions: Identifying high-dimensional
  multilinear systems via compressed sensing,'' in \emph{Proc. Allerton
  Conference Commun., Control, and Comput.}, Sept. 2010, pp. 1589--1596.

\bibitem{VAMP2017}
S.~Rangan, P.~Schniter, and A.~K. Fletcher, ``Vector approximate message
  passing,'' in \emph{Proc. Int. Symp. Inf. Theory (ISIT)}, July 2017, pp.
  1588--1592.

\bibitem{Tibshirani1996}
R.~Tibshirani, ``Regression shrinkage and selection via the {LASSO},'' \emph{J.
  Royal Stat. Soc. Series B (Methodological)}, vol.~58, no.~1, pp. 267--288,
  1996.

\bibitem{VilaSchniter2011}
J.~Vila and P.~Schniter, ``Expectation-maximization {B}ernoulli-{G}aussian
  approximate message passing,'' in \emph{Proc. IEEE 45th Asilomar Conf.
  Signals, Syst., and Comput.}, Nov. 2011, pp. 799--803.

\bibitem{Simoncelli1996}
E.~P. Simoncelli and E.~H. Adelson, ``Noise removal via {B}ayesian wavelet
  coring,'' in \emph{Proc. Int. Conf. Image Process.}, vol.~1.\hskip 1em plus
  0.5em minus 0.4em\relax IEEE, Sept. 1996, pp. 379--382.

\bibitem{Gauss1809theoria}
C.~F. Gauss, \emph{{Theoria motus corporum coelestium in sectionibus conicis
  solem ambientium}}, 1809.

\end{thebibliography}

\end{document}